\documentstyle[twocolumn,latex8,times,epsf]{article}

\newtheorem{theorem}{Theorem}[section]

\newcommand{\qed}{\hfill $\Box$}
\newenvironment{proof}{{\bf Proof:}}{\qed \par}

\newcommand{\mplus}{m^{+}(\beta)}
\newcommand{\rplus}{r^{+}}

\newcommand{\Fout}{F^{out}}
\newcommand{\Fin}{F^{in}}
\newcommand{\coupon}{R_{\beta,t}}
\newcommand{\coupona}[1]{R_{#1,t}}

\newcommand{\ignore}[1]{}

\def\QED{\hfill \qed\lower 7pt\null\par}

\begin{document}
\pagestyle{empty}
\date{}

\title{On Diffusing Updates in a Byzantine Environment}

\author{
Dahlia Malkhi\thanks{AT\&T Labs Research, Florham Park, NJ, USA, and Institute of Computer
Science, The Hebrew University of Jerusalem, Israel; {\sf dalia@cs.huji.ac.il}}
\and
Yishay Mansour\thanks{AT\&T Labs Research, Florham Park, NJ, USA, and Computer Science
Department, Tel Aviv University, Israel; {\sf mansour@math.tau.ac.il}}
\and
Michael K.\ Reiter\thanks{Bell Laboratories, Murray Hill, NJ, USA; {\sf
reiter@research.bell-labs.com}}}
\maketitle
\thispagestyle{empty}

\subsection*{\centering Abstract}
{ \em We study how to efficiently diffuse updates to a large distributed
system of data replicas, some of which may exhibit arbitrary
(Byzantine) failures.  We assume that strictly fewer than $t$ replicas
fail, and that each update is initially received by at least $t$
correct replicas.  The goal is to diffuse each update to all correct
replicas while ensuring that correct replicas accept no updates
generated spuriously by faulty replicas.  To achieve reliable
diffusion, each correct replica accepts an update only after receiving
it from at least $t$ others.  We provide the first analysis of
epidemic-style protocols for such environments.  This analysis is
fundamentally different from known analyses for the benign case due to
our treatment of fully Byzantine failures---which, among other things,
precludes the use of digital signatures for authenticating forwarded
updates.  We propose two epidemic-style diffusion algorithms and two
measures that characterize the efficiency of diffusion algorithms in
general.  We characterize both of our algorithms according to these
measures, and also prove lower bounds with regards to these measures
that show that our algorithms are close to optimal.
}

\section{Introduction}

A diffusion protocol is the means by which an update initially known
to a portion of a distributed system is propagated to the rest of the
system.  Diffusion is useful for driving replicated data toward a
consistent state over time, and has found application for this
purpose, e.g., in USENET News~\cite{LOM94}, and in the
Grapevine~\cite{BLNS82} and Clearinghouse~\cite{OD81} systems.  The
quality of a diffusion protocol is typically defined by the delay
until the update has reached all replicas, and the amount of message
traffic that the protocol generates.

In this paper, we provide the first study of update diffusion in
distributed systems where components can suffer Byzantine failures.
The framework for our study is a network of data replicas, of which
strictly less than some threshold $t$ can fail arbitrarily, and to
which updates are introduced continually over time.  For example,
these updates may be sensor readings of some data source that is
sampled by replicas, or data that the source actively pushes to
replicas.  However, each update is initially received only by a subset
of the correct replicas of some size $\alpha \ge t$, and so replicas
engage in a diffusion protocol to propagate updates to all correct
replicas over time.  Byzantine failures impact our study in that a
replica that does not obtain the update directly from the source must
receive copies of the update from at least $t$ different replicas
before it ``accepts'' the update as one actually generated by the
source (as opposed to one generated spuriously by a faulty replica).

In our study, we allow fully Byzantine failures, and
thus cannot rely on digital signatures to authenticate the original
source of a message that one replica forwards to others.  While
maximizing the fault models to which our upper bounds apply, avoiding
digital signatures also strengthens our results in other respects.
First, in a network that is believed to intrinsically provide the
correct sender address for each message due to the presumed difficulty
of forging that address, avoiding digital signatures avoids the
administrative overheads associated with distributing cryptographic
keys.  Second, even when the sender of a message is not reliably
provided by the network, the sender can be authenticated using
techniques that require no cryptographic assumptions (for a survey of
these techniques, see~\cite{Sim92}).  Employing digital signatures, on
the other hand, would require assumptions limiting the computational
power of faulty replicas.  Third, pairwise authentication typically
incurs a low computation overhead on replicas, whereas digitally
signing each message would impose a significantly higher overhead.

To achieve efficient diffusion in our framework, we suggest two
round-based algorithms: ``Random'', which is an epidemic-style
protocol in which
each replica sends messages to randomly chosen replicas in each
round, and ``$\ell$-Tree-Random'', which diffuses updates along a tree
structure.  For these algorithms, two measures of quality are studied:
The first one, {\em delay}, is the expected number of rounds until any
individual update is accepted by all correct replicas in the
system. The delay measure expresses the speed of propagation. The
second, {\em fan-in}, is the expected maximum number of messages
received by any replica in any round from correct replicas. Fan-in is
a measure of the load inflicted on individual replicas in the common
case, and hence, of any potential bottlenecks in execution.  We
evaluate these measures for each of the protocols we present.  In
addition to these results, we prove a lower bound of
$\Omega(\frac{t}{\Fout}\log{\frac{n}{\alpha}})$ on the delay of any
diffusion protocol, where $\Fout$ is the ``fan-out'' of the protocol,
i.e., a bound on the number of messages sent by any correct process in
any round.  We also show an inherent tradeoff between good (low)
latency and good (low) fan-in, namely that their product is at least
$\Omega(tn/\alpha)$.  Using this tradeoff, we demonstrate that our
protocols cover much of the spectrum of optimal-delay protocols for
their respective fan-in to within logarithmic factors.

\begin{figure}[htb]
\begin{center}
\hspace*{-0.2in}\epsffile{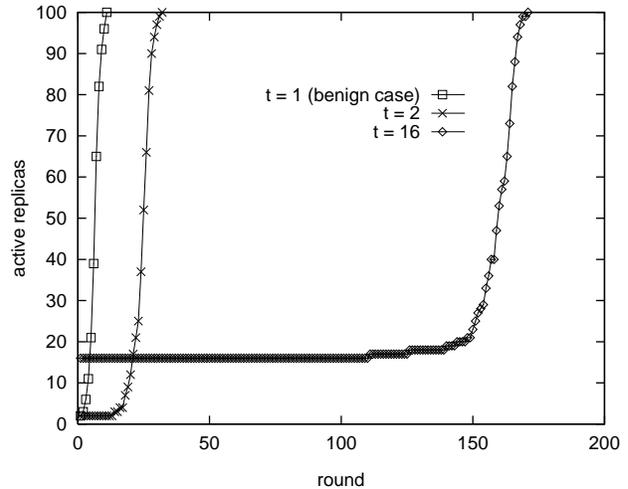}
\caption{Delay of random propagation with $n=100$, $\alpha=t+1$}
\end{center}
\label{fig:diffusion}
\end{figure}

We emphasize that our treatment of full Byzantine failures renders our
problem fundamentally different from the case of crash failures only.
Intuitively, any diffusion process has two phases: In the first phase,
the initially active replicas for an update send this update, while
the other replicas remain inactive. This phase continues while
inactive replicas have fewer than $t$ messages. In the second phase,
new replicas become active and propagate updates themselves, resulting
in an exponential growth of the set of active replicas. In
Figure~\ref{fig:diffusion} we depict the progress of epidemic
diffusion. The figure shows the number of active replicas plotted
against round number, for a system of $n = 100$ replicas with
different values of $t$, where $\alpha = t+1$.  The case $t=1$ is
indistinguishable from diffusion with benign failures only, since a
single update received by a replica immediately turns it into an
active one.  Thus, in this case, the first phase is degenerate, and
the exponential-growth phase occurs from the start. Previous work has
analyzed the diffusion process in that case, proving
propagation delay~\cite{DGH+87} that is logarithmic in the number of
replicas.  However, in the case that we
consider here, i.e., $t \ge 2$, the delay is dominated by the initial
phase.

The rest of the paper is organized as follows. In
Section~\ref{sec:motivation} we illustrate specific applications for
which Byzantine message diffusion is suitable, and which motivated our
study. We discuss related work in Section~\ref{sec:related}.  In
Section~\ref{sec:model} we lay out assumptions and notation used
throughout the paper, and in Section~\ref{sec:measures} we define our
measures of diffusion performance.  In Section~\ref{sec:general} we
provide general theorems regarding the delay and fan-in of diffusion
protocols.  In Section~\ref{sec:random} we introduce our first
diffusion protocol, Random, and analyze its properties, and in
Section~\ref{sec:tree} we describe the $\ell$-Tree-Random protocol and
its properties. We summarize and discuss our results in
Section~\ref{sec:discussion}. Section~\ref{sec:simulation} provides
simulation results that demonstrate the likely behavior of our
protocols in practice.
We conclude in Section~\ref{sec:conclusion}.

\subsection{Motivation}
\label{sec:motivation}

The motivating application of our work on message diffusion is a data
replication system called Fleet.  (Fleet is not yet documented, but is
based on similar design principles as a predecessor system called
Phalanx~\cite{MR98b}.)  Fleet replicates data so that it will
survive even the malicious corruption of some data replicas, and does
so using adaptations of quorum systems to such
environments~\cite{MR98a}.  A characteristic of these replication
techniques that is important for this discussion is that each update
is sent to only a relatively small subset (quorum) of servers, but one
that is guaranteed to include $t$ correct ones, where the number of
faulty replicas is assumed to be less than $t$.  Thus, after an
update, most correct replicas have not actually received this update,
and indeed any given correct replica can be arbitrarily out-of-date.

While this local inconsistency does not impact the global consistency
properties of the data when the network is connected (due to the
properties of the quorum systems we employ), it does make the system
more sensitive to network partitions.  That is, when the network
partitions---and thus either global data consistency or progress of
data operations must be sacrificed---the application may dictate that
data operations continue locally even at the risk of using stale data.
To limit how stale local data is when the network partitions, we use a
diffusion protocol while the network is connected to propagate updates
to all replicas, in the background and without imposing additional
overhead on the critical path of data operations.  In this way, the
system can still efficiently guarantee strict consistency in case a
full quorum is accessed, but can additionally provide relaxed
consistency guarantees when only local information is used.

Another variation on quorum systems, {\em probabilistic quorum
systems}~\cite{MRW97,MRWW98}, stands to benefit from properly designed
message diffusion in different ways than above. Probabilistic quorum
systems are a means for gaining dramatically in performance and
resilience over traditional (strict) quorum systems by allowing a
marginal, controllable probability of inconsistency for data reads.
When coupled with an effective diffusion technique, the
probability of inconsistency can be driven toward zero when updates
are sufficiently dispersed in time.

More generally, diffusion is a fundamental mechanism for driving
replicated data to a consistent state in a highly decentralized
system. Our study sheds light on the use of diffusion protocols in
systems where arbitrary failures are a concern, and may form a basis
of solutions for disseminating critical information in survivable
systems (e.g., routing table updates in a survivable network
architecture).

\subsection{Related work}
\label{sec:related}

The style of update diffusion studied here has previously been studied
in systems that can suffer benign failures only.  Notably, Demers
et.al.~\cite{DGH+87} performed a detailed study of epidemic algorithms
for the benign setting, in which each update is initially known at a
single replica and must be diffused to all replicas with minimal
traffic overhead.  One of the algorithms they studied, called {\em
anti-entropy} and apparently initially proposed in~\cite{BLNS82}, was
adopted in Xerox's Clearinghouse project (see~\cite{DGH+87}) and the
Ensemble system~\cite{BHOXBM98}.  Similar ideas also underly
IP-Multicast~\cite{Dee89} and MUSE (for USENET News
propagation)~\cite{LOM94}.  This anti-entropy technique forms the
basis for one of the algorithms (Random) that we study here.  As
described previously, however, the analysis provided here of
the epidemic-style update diffusion is fundamentally different for
Byzantine environments than for environments that suffer benign
failures only.

Prior studies of update diffusion in distributed systems that can
suffer Byzantine failures have focused on single-source broadcast
protocols that provide reliable communication to replicas and replica
agreement on the broadcast value (e.g.,~\cite{LSP82,DS83,BT85,MR96}),
sometimes with additional ordering guarantees on the delivery of
updates from different sources (e.g.,~\cite{Rei94,CASD95,MM95,KMM98}).
The problem that we consider here is different from these works in the
following ways.  First, in these prior works, it is assumed that one
replica begins with each update, and that this replica may be
faulty---in which case the correct replicas can agree on an arbitrary
update.  In contrast, in our scenario we assume that at least a
threshold $t > 1$ of {\em correct} replicas begin with each update,
and that only these updates (and no arbitrary ones) can be accepted by
correct replicas.  Second, these prior works focus on certain
reliability, i.e., guaranteeing that all correct replicas (or all
correct replicas in some agreed-upon subset of replicas) receive the
update.  Our protocols diffuse each update to all correct servers only
with some probability that is determined by the number of rounds for
which the update is propagated before it is discarded.  Our goal is to
analyze the number of rounds until the update is expected to be
diffused globally and the load imposed on each replica as measured by
the number of messages it receives in each round.

\section{System model}
\label{sec:model}

We assume a system of $n$ replicas, denoted $p_1, \ldots, p_n$.  A
replica that conforms to its I/O and timing specifications is said
to be {\em correct}.  A {\em faulty} replica is one that deviates from
its specification.  A faulty replica can exhibit arbitrary behavior
(Byzantine failures).  We assume that strictly fewer than $t$ replicas
fail, where $t$ is a globally known system parameter.

Replicas can communicate via a completely connected point-to-point
network.  Communication channels between correct replicas are reliable
and authenticated, in the sense that a correct replica $p_i$ receives
a message on the communication channel from another correct replica
$p_j$ if and only if $p_j$ sent that message to $p_i$.  Moreover, we
assume that communication channels between correct replicas impose a
bounded latency $\Delta$ on message transmission; i.e., communication
channels are {\em synchronous}.  Our protocols will also work
to diffuse updates in an asynchronous system, but in this case we can
provide no delay or fan-in analysis.  Thus, we restrict our attention
to synchronous systems here.

Our diffusion protocols proceed in synchronous rounds.  A system
parameter, {\em fan-out}, denoted $\Fout$,
bounds from above the number of messages any
correct replica sends in a single round.
A replica receives and processes all messages sent to
it in a round, before the next round starts.
Thus, rounds begin at least $\Delta$ time units apart.

Each update $u$ is introduced into the system at a set $I_u$ of
$\alpha \ge t$ {\em correct} replicas, and possibly also at some
other, faulty replicas.  We assume that all replicas in $I_u$
initially receive $u$ simultaneously (i.e., in the same round).  The
goal of a diffusion protocol is to cause $u$ to be {\em accepted} at
all correct replicas in the system.  The update $u$ is accepted at
correct replica $p_i$ if $p_i \in I_u$ or $p_i$ has received $u$ from
$t$ other distinct replicas.  If $p_i$ has accepted $u$, then we also
say that $p_i$ is {\em active} for $u$ (and is {\em passive}
otherwise). In all of our diffusion protocols, we assume that each
message contains all the updates known to the sender, though in
practice, obvious techniques can reduce the actual number of updates
sent to necessary ones only.

\section{Measures}
\label{sec:measures}

We study two complexity measures: {\em delay\/} and {\em fan-in}.  For
each update, the delay is the expected number of rounds from the time the
update is introduced to the system until all correct replicas accept
the update.
Formally, let $\eta_u$ be the round number in which update $u$ is introduced
to the system, and let $\tau_p^u$ be the round in which
a correct replica $p$ accepts update $u$.
The delay is $E[\max_p \{ \tau_p^u \}-\eta_u]$,
where the expectation is over the random choices of the algorithm
and the maximization is over correct replicas.

We define Fan-in to be the expected maximum number of messages that
any correct replica receives in a single round from correct replicas
under all possible failure scenarios.  Formally, let $\rho_p^i$ be the
number of messages received in round $i$ by replica $p$ from correct
replicas.  Then the fan-in in round $i$ is $E[\max_{p,C}\{\rho_p^i\}]$,
where the maximum is taken with respect to all correct replicas $p$
and all failure configurations $C$ containing fewer than $t$ failures.
An amortized fan-in is the
expected maximum number of messages received over multiple rounds,
normalized by the number of rounds.  Formally, a $k$-amortized fan-in
starting at round $l$ is $E[\max_{p,C}\{\sum_{i=l}^{l+k} \rho_p^i/k\}]$.
We emphasize that fan-in and amortized fan-in are measures only for
messages from correct replicas.  Let $\Fin$ denotes the fan-in.
In a round a correct replica may receive
messages from $\Fin+t-1$ different replicas, and may receive any number
of messages from faulty replicas.

A possible alternative is to define fan-in as an absolute bound
limiting the number of replicas from which each correct replica will
accept messages in each round.  However, this would render the system
vulnerable to ``denial of service'' attacks by faulty replicas: by
sending many messages, faulty replicas could force messages from
correct replicas to compete with up to $t-1$ messages from faulty
replicas in every round, thus significantly changing the behavior of
our protocols.

\section{General Results}
\label{sec:general}

In this section we present general results concerning the delay and
fan-in of any propagation algorithm. Our first result is a lower bound
on delay, that stems from the restriction on fan-out, $\Fout$.  This
lower bound is for the worst case delay, i.e., when faulty replicas
send no messages.

\begin{theorem}
\label{prop:lbdelay}
The delay of any diffusion algorithm $A$ is $\Omega(\frac{t}{\Fout}
\log{\frac{n}{\alpha}})$.
\end{theorem}

\begin{proof}
Let $u$ be any update, and let $m_k$ denote the total number of times
$u$ is sent by correct processes in rounds $\eta_u + 1, \ldots, \eta_u
+ k$ in $A$.  Denote by $\alpha_k$ the number of correct replicas that
have accepted update $u$ by the time round $\eta_u + k$ completes.
Since $t$ copies of update $u$ need to reach a replica (not in $I_u$)
in order for it to accept the update, we have that $\alpha_k \leq
\alpha+m_k/t$.  Furthermore, since at most $\Fout \alpha_k$ new
updates are sent by correct processes in round $\eta_u + k+1$, we have
that $m_{k+1} \le m_{k} + \Fout \alpha_{k} \le \Fout \sum_{j=0}^k
\alpha_j$, where $\alpha_0=\alpha$.  By induction on $k$, it can be
shown that $\alpha_k \le \alpha (1 + \frac{\Fout}{t})^k$.  Therefore,
for $k < \frac{t}{\Fout} \log{\frac{n}{\alpha}}$ we have that
$\alpha_k < n$, which implies that not all the replicas are active for
update $u$.
\end{proof}

\vspace{0.1in}
\noindent
The next theorem shows that there is an inherent tradeoff
between fan-in and delay.

\begin{theorem}
\label{prop:tradeoff}
Let $A$ be any propagation algorithm. Denote by $D$ its delay,
and by $\Fin$ its $D$-amortized fan-in.
Then $D\Fin = \Omega(tn/\alpha)$, for $t \ge 2\log{n}$.
\end{theorem}

\begin{proof}
Let $u$ be any update.
Since the $D$-amortized fan-in of $A$ is $\Fin$, with
probability $0.9$ (where $0.9$ is arbitrarily chosen here as some
constant between $0$ and $1$), the number of messages received
(from correct replicas) by any replica in rounds $\eta_u+1,
\ldots, \eta_u+D$ is less than $10D\Fin$.  From now on we will
assume that every replica $p_j$ receives at most $10D\Fin$ messages in
rounds $\eta_u+1,\ldots,\eta_u+D$.  This means that for each $p_j$, if
$p_j$ is updated by a set $S_j$ of replicas during rounds $\eta_u + 1,
\ldots, \eta_u + D$, then $|S_j|\leq 10D\Fin$.
Some replica $p_j$ becomes active for $u$ if out of the
updates in $S_j$ at least $t$ are from $I_u$,
i.e. $|S_j\cap I_u| \geq t$.
In order to show the lower bound, we need to exhibit an initial set
$I_u$, such that if $10D\Fin$ is too small then no replica becomes
active. More specifically,
for $D \le \frac{1}{2} \frac{nt}{10\Fin \alpha}$,
we show that there exists a set $I_u$ such that for each
$p_j$, we have $|S_j\cap I_u| < t$.

We choose the initial set $I_u$ as a random subset of $\{p_1, \ldots,
p_n\}$ of size $\alpha$.  Let $X_j$ denote the number of replicas in $I_u$
from which messages are received by replica $p_j$ during rounds $\eta_u+1,
\ldots,\eta_u+D$,
i.e.,  $X_j=|S_j\cap I_u|$.
Since $p_j$ receives at most $10D\Fin$ messages in these rounds, we get
\begin{eqnarray*}
Prob[X_j \geq k] &<&
  \sum_{i=k}^{10D\Fin} \frac{{10D\Fin \choose i} {{n-10D\Fin}
  \choose {\alpha-i}}}{{n \choose \alpha}} \\
  &<& \sum_{i=k}^{n} {10D\Fin \choose i} \left(\frac{\alpha}{n}\right)^i \\
  &\leq& \left(\frac{10e D\Fin\alpha}{kn}\right)^k~c~,
\end{eqnarray*}
where the constant $c$ is at most $2$ if $D \le \frac{1}{2}
\frac{nk}{10e\Fin \alpha}$, and hence we have that $Prob[X_j \geq t] <
(1/2)^t$. By our assumption that $t\geq 2\log n$, we have that
$Prob[X_j \geq t] < 1/n^2$.  This implies that the probability that
all the $X_j$ are at most $t$ is at least $1-(1/n)$.

We have shown that for most subsets $I_u$ if
$D \le \frac{1}{2} \frac{nt}{10e\Fin \alpha}$
no new replica would become active.
Therefore, for some specific $I_u$ it also holds.
(In fact it holds for most.)

Recall that at the start of the proof we assumed that the
$D$-amortized fan-in is at most $10\Fin$. This holds with probability at
least $0.9$.  Therefore in $0.9$ of the runs the delay is at least
$\frac{1}{2} \frac{nt}{10e\Fin \alpha}$, which implies that the expected
delay is $\Omega(\frac{nt}{\Fin \alpha})$.
\end{proof}

\section{Random Propagation}
\label{sec:random}

In this section, we present a random diffusion method
and examine its delay and fan-in measures.
In this algorithm, which we refer to as simply ``Random'', each
replica, at each round, chooses $\Fout$ replicas uniformly at random from all
replicas and sends messages to them.  This method is similar to the
``anti-entropy'' method of~\cite{BLNS82,DGH+87}.

In the next theorem we use the notation
of
\[
\coupon=\beta \sum_{j=\beta-t+1}^{\beta} 1/j \approx \beta
\log\frac{\beta}{\beta-t+1} + O\left(\frac{\beta}{\beta-t+1}\right),
\]
which is the result of
the analysis of the coupon collector problem, i.e., the expected
number of steps for collecting $t$ distinct `coupons' out of $\beta$
different ones by random polling (see~\cite[ch.\ 3]{MR95}).
It is worth discussing how $\coupon$ behaves for various values
of $\beta$ and $t$. For $\beta=t$ we have $\coupon \approx t\log t$.
For $\beta\geq 2t$ we have $\coupon \leq 1.5 t$. For all $\beta \ge
t$, we have $\coupon \ge t$.
This implies that if the initial set size $\beta$ is very close to $t$,
then we have a slightly superlinear behavior of $\coupon$
as a function of $t$, while if $\beta$ is a fraction away from $t$
then we have $\coupon$ as a linear function in $t$.

\begin{theorem}
\label{prop:epdelay}
The delay of the Random algorithm is
$O\left( \frac{\coupona{\alpha}}{\Fout}
(\frac{n}{\alpha})^{(1-\frac{1}{2\coupona{\alpha}})} + \frac{\log(n)}{\Fout}
\right)$
for $2 < t \le n/4$.
\end{theorem}

\begin{proof}
The outline of the proof is as follows. For the most part, we consider
bounds on the number of messages sent, rather than directly on the
number of rounds.  It is more convenient to argue about the number of
messages, since the distribution of the destination of each replica's
next message is fixed, namely uniform over all replicas.  As long as
we know that there are between $\alpha$ and $2\alpha$ replicas active
for $u$, we can translate an upper bound on the number of messages
to an approximate upper bound on the number of rounds.

More specifically, so long as the number $\beta$ of active replicas
does not reach a quarter of the system, i.e., $\alpha \le \beta \le
n/4$, we study $\mplus$, an upper bound on the number of messages
needed to be sent such that with high probability, $1-q^{+}(\beta)$,
we have $\beta$ new replicas change state to active.  We then analyze
the algorithm as composed of phases starting with $\beta = 2^j
\alpha$.  The upper bound on the number of messages to reach half the
system is $\sum_{j=0}^{\ell} m^{+}(2^j \alpha)$, the bound on the
number of rounds is
$\sum_{j=0}^{\ell} m^{+}(2^j \alpha)/(2^j\Fout\alpha)$,
and the error probability is at most $\sum_{j=0}^{\ell} q^{+}(2^j \alpha)$,
where $\ell=\log(n/2\alpha)-1$.  In the analysis we assume for
simplicity that $n=2^j\alpha$ for some $j$, and this implies that in
the last component we study, there are at most $n/4$ active replicas.

At the end, we consider the case where $\beta > n/4$, and
bound from above the number of rounds needed to complete the
propagation algorithm. This case adds only an additive factor of
$O((t+\log n)/\Fout)$
to the total delay.

We start with the analysis of the number of messages required to move
from $\beta$ active replicas to $2\beta$, where $\beta \le n/4$.
For any $m$, let $N^m_i$ be the number of messages that $p_i$ received,
out of the first $m$ messages, and let
$S^m_i$ be the number of distinct replicas that sent the $N^m_i$
messages.
Let $U^m_i$ be an indicator variable such that $U^m_i =
1$ if $p_i$ receives messages from $t$ or more
distinct replicas after $m$ messages are sent, and $U^m_i = 0$ otherwise.
I.e. $U^m_i=1$ if and only if $S^m_i \geq t$.

We now use the {\em coupon collector's} analysis
to bound the probability that $S^m_i \geq t$ when $N^m_i$ messages are
received. Thus, a replica needs to get an expected $\coupon$ messages
before $S^m_i \geq t$, and so with probability $\le 1/2$ it would need
more than $2\coupon$ messages to collect $t$ different messages.  For
$m \le n+2\coupon$ we have that
\begin{eqnarray*}
\lefteqn{Prob[U^m_i=1]} \\
  & \ge & Prob[N^m_i = 2\coupon] Prob[ U^m_i = 1| N^m_i = 2\coupon] \\
  & \ge & {m \choose {2\coupon}} \left(\frac{1}{n}\right)^{2\coupon}
          \left(1-\frac{1}{n}\right)^{m-2\coupon} \left(\frac{1}{2}\right) \\
  & \ge & \left(\frac{m}{2\coupon}\right)^{2\coupon}
          \left(\frac{1}{n}\right)^{2\coupon}
          e^{-(m-2\coupon)/n} \left(\frac{1}{2}\right) \\
  & \ge & \left(\frac{m}{ 2n \coupon}\right)^{2\coupon}
          \left(\frac{1}{6}\right)
\end{eqnarray*}

Let $U^m$ denote the number of replicas that received messages
from $t$ or more
replicas after $m$ messages are sent, i.e., $U^m =\sum_{i=\beta+1}^{n} U^m_i$,
where the active replicas are $p_1, \ldots p_\beta$.
For $\beta \le n/4$  we have,
\begin{eqnarray*}
  E[U^m]
& \ge &
  (n-\beta)
  \left(\frac{m}{2 n\coupon}\right)^{2\coupon}
  \left(\frac{1}{6}\right) \\
& \ge &
  \frac{n}{12}
  \left(\frac{m}{2 n\coupon}\right)^{2\coupon} ,
\end{eqnarray*}

where the right inequality uses the fact that $\beta \le n/4$.

Our aim is to analyze the distribution of $U^m$. More specifically, we would
like to find $\mplus$ such that,
\[
Prob[U^m \geq 2\beta] > 1-q^{+}(\beta)
\]
for any $m>\mplus$.

Generally, the analysis is simpler when the random variables are
independent.  Unfortunately, the random variables $U^m_i$ are not
independent, but using a classical result by Hoeffding \cite[Theorem
4]{Hoe63}, the dependency works only in our favor.  Namely, let
$X_i^m$ be i.i.d.\ binary random variables with
$Prob[X_i^m=1]=Prob[U_i^m=1]$, and $X^m=\sum_{i=1}^n X_i^m$.  Then,
\[
Prob[U^m-E[U^m]\ge \gamma] \le Prob[X^m -E[X^m
]\ge \gamma] ~.
\]
 From now on we will prove the bounds for $X^m$ and they will apply
also to $U^m$. First, using a Chernoff bound (see~\cite{KV94}) we have that,

\[
Prob\left[X^{\mplus} \le \frac{1}{2} E[X^{\mplus}] \right] \le
e^{-\frac{E[X^{\mplus}]}{8}} ~.
\]
\noindent
For
  $\mplus=2 n\coupon (24 \beta/n)^{1/2\coupon}$,
we have
$E[X^{\mplus}]\geq 2\beta$, and hence
\[
Prob[X^{\mplus} \le \beta ] \le e^{-\beta/4} = q^{+}(\beta) ~.
\]

For the analysis of the Random algorithm, we view the algorithm as
running in phases so long as $\beta \le n/4$.  There will be
$\ell=\log(n/2\alpha)-1$ phases, and in each phase we start with
$\beta = 2^j \alpha$ initial replicas, for $0\leq j\leq \ell$.  The
$j$th phase runs for $m^{+}( 2^j \alpha)/(\Fout 2^j \alpha)$ rounds.
We say that a phase
is ``good'' if by the end of the phase the number of active replicas has
at least doubled.  The probability that some phase is not good is
bounded by,
\[
\sum_{j=0}^{\ell} q^{+}(2^j \alpha) =
\left(\sum_{j=0}^{\ell} e^{-2^j \alpha/4}\right)
\leq 2e^{-\alpha/4}\leq 1/2,
\]
for $\alpha\geq 6$.  Assuming that all the phases are good, at the end
half of the replicas are active.

The number of rounds until half the system is active is at most,
\begin{eqnarray*}
\sum_{j=0}^{\ell} \frac{m^{+}(2^j \alpha)}{\Fout 2^j \alpha} & = &
\sum_{j=0}^{\ell} \frac{ 2 n\coupona{2^j\alpha}(24 \times 2^j
\alpha/n)^{1/(2\coupona{2^j\alpha})}}{\Fout 2^j \alpha} \\
& \leq &
\frac{ 2n \coupona{\alpha}}{\Fout\alpha} \sum_{j=0}^{\ell}
\frac{ (24 \cdot 2^j
\alpha/n)^{1/(2\coupona{\alpha})}}{ 2^j } \\
& = &
 O \left(\frac{\coupona{\alpha}}{\Fout}
 \left(\frac{n}{\alpha}\right)^{1-\frac{1}{2\coupona{\alpha}}}\right),
\end{eqnarray*}
where we used here the fact that $\coupon$ is a decreasing function in
$\beta$.

\ignore{
We now consider two cases, $\alpha \ge 2t$ and $\alpha < 2t$. In the
first case, when $\alpha \ge 2t$,
the number of rounds until half the system is active is,
\begin{eqnarray}
\sum_{j=0}^{\ell} \frac{m^{+}(2^j \alpha)}{\Fout 2^j \alpha} & = &
\sum_{j=0}^{\ell} \frac{ n 2\coupona{2^j\alpha}(24 \times 2^j
\alpha/n)^{1/(2\coupona{2^j\alpha})}}{\Fout 2^j \alpha} \\
& \le &
\sum_{j=0}^{\ell} \frac{ n 2 \times 1.5t \times 24^\frac{1}{2t}
(2^j\alpha/n)^\frac{1}{2\times 1.5t}}{\Fout 2^j \alpha} \\
& = &
 \frac{2\times 1.5t \times 24^\frac{1}{2t}}{\Fout}
\left(\frac{n}{\alpha}\right)^{1-\frac{1}{3t}}
\left[       \sum_{j=0}^{\log(n/2\alpha) - 1} (1/
2^j)^{1-\frac{1}{3t}} \right]\\
& = &
 O(\frac{t}{\Fout}(\frac{n}{\alpha})^{1-\frac{1}{3t}}),
\end{eqnarray}

In the second case, when $\alpha < 2t$, the number of round is, moving
from $\alpha$ to $2\alpha$ replicas may require $(n 2 t\log{t}
(24\alpha/n)^{1/2t\log{t}})/(\Fout \alpha)$ rounds, and so the total
number of rounds is
\[
 O(\frac{t\log{t}}{\Fout}(\frac{n}{\alpha})^{1-\frac{1}{2t\log{t}}}) ,
\]
}

We now reach the last stage of the algorithm, when $\beta \ge
n/2$. Unfortunately, there are too few passive replicas to use the
analysis above for $\mplus$, since we cannot drive the expectation
of $X^{m}$ any higher than $\beta$.  We therefore employ a different
technique here.

We give an upper bound on the expected number of rounds for
completion at the last stage.  Fix any replica $p$, and let $V_i$ be
the number of new update in round $i$ that $p$ receives.
Since $t \le n/4$, we have $\beta - t \ge n/4$, and so:
\[
E[V_i] = (\beta-t)\frac{\Fout}{n} \geq \frac{\Fout}{4}.
\]
Let $V^r$ denote the number of new updates received by $p$ in $r$
rounds, hence $V^r = \sum_{i=1}^{r} V_i$.  Then, $E[V^r] \geq r\Fout/4$.
Using the Chernoff bound we have,
\[
        Prob[V^r < r\Fout/8 ] \le e^{- \Fout  r/64}
\]
Let $\rplus = (8t+128\log(n))/\Fout$.
The probability that $V^{\rplus}$ is less than $t$
is at most $1/n^2$.
The probability that some replica receives less than $t$ new updates
in $\rplus$ rounds is thus less than $1/n$, and so in an expected $O((t
+ \log(n))/\Fout)$ rounds the algorithm terminates.

\ignore{ 

We give an upper bound on the expected number of rounds for
completion at the last stage.  Fix any replica $p$, and let $V_i$ be
an indicator variable such that $V_i = 1$ iff $p$ receives a new
update in round $i$, and $V_i = 0$ otherwise.
Since $t \le n/4$, $\beta - t \ge n/4$, and so:
\[
 Prob[V_i = 0] \le
        \left(1 - \frac{1}{n}\right)^{(\beta-t)} \le
        \left(1 - \frac{1}{n}\right)^{n/4} \le
        e^{-{1/4}}~.
\]
Let $V^r$ denote the number of new updates received by $p$ in $r$
rounds, hence $V^r = \sum_{i=1}^{r} V_i$.  Let
$b = 1-e^{-1/4}$. Thus,
\[
        br \le E[V^r] ~.
\]

\noindent
Using the Chernoff bound we have that for any $r$,
\[
        Prob[V^r < br/2 ] \le e^{-b  r/8}
\]
Let $\rplus = (2t+16\log(n))/b$.
The probability that $V^{\rplus}$ is less than $t$
is at most $1/n^2$.
The probability that some replica receives less than $t$ new updates
in $\rplus$ rounds is thus less than $1/n$, and so in an expected $O(t
+ \log(n))$ rounds the algorithm terminates.
}

Putting the two bounds together, we have an expected
$O(\frac{\coupona{\alpha}}{\Fout}
\left(\frac{n}{\alpha}\right)^{1-\frac{1}{2\coupona{\alpha}}}
+ \frac{\log(n)}{\Fout})$ number of rounds.
\end{proof}
\vspace{0.1in}\mbox{}\\

The proof of the theorem reveals that it takes the same order of
magnitude of rounds just to add $\alpha$ more replicas to be active as
it is to make all the replicas active.  This is due to the phenomena
that having more replicas active reduces the time to propagate the
update.  This is why we have a rapid transition from not having
accepted the update by any new replicas to having them accepted by all
replicas.

Note that when $t=\Omega(\log n)$, then simply by sending to replicas
in a round-robin fashion, the initially active replicas can propagate
an update in $O(\frac{nt}{\alpha\Fout})$ rounds to the rest of the
system. The Random algorithm reaches essentially the same bound in
this case. This implies that the same delay would have been reached if
the replicas that accepted the update would not have participated in
propagating it (and only the original set of replicas would do all the
propagating).
Finally, note that in failure-free runs of the system, the upper bound
proved in Theorem~\ref{prop:epdelay} is also the lower bound on the
expected delay, i.e., it is tight.

The next theorem bounds the fan-in of the random algorithm.
Recall that the fan-in measure is with respect to the messages sent
by the correct replicas.

\begin{theorem}
\label{prop:epfan-in}
The fan-in of the Random algorithm is $O(\Fout + \log{n})$,
and when $\Fout \le 1/4\log n$, it is $O(\frac{\Fout +
\log{n}}{\log{\log{n}} - \log{\Fout}})$.
(Note that when $\Fout=1$, this fan-in is
$O(\frac{\log{n}}{\log{\log{n}}})$).
The $(\log n)$-amortized fan-in is $O(\Fout)$.
%
\end{theorem}

\begin{proof}
The probability that a replica receives $k$ messages or more in one
round is bounded by ${n \Fout \choose k } (1/n)^k$, which is bounded by
$(e\Fout /k)^k$. For $k = c(\Fout + \log{n})$,
this bound is
$O(1/n^2)$, for some $c>0$.
Hence the probability that any replica
receives more than $k = c(\Fout + \log{n})$ in a round is small.
Therefore, the fan-in is bounded by $O(\Fout +\log n)$.
If $\Fout \le 1/4 \log n$, then for $k = c
\frac{\Fout+\log{n}}{\log{\log{n}}-\log{\Fout}}$,
this bound is $O(1/n^2)$, for some $c>0$.
Therefore, in this case, the fan-in is bounded by
$O(\frac{\Fout+\log{n}}{\log{\log{n}}-\log{\Fout}})$.

The probability that in $\log n$ rounds a specific replica receives
more than $k=6\Fout  \log{n}$
messages is bounded by ${{n\Fout\log n} \choose {k}}
(1/n)^{k}$ which is bounded by $1/n^2$.  The probability that any
replica receives more than $k=6\Fout \log n$ messages is bounded by
$1/n$. Thus, the $(\log n)$-amortized fan-in is at most $O(\Fout)$.
\end{proof}

\section{Tree-Random}
\label{sec:tree}

The Random algorithm above is one way to propagate an update.  Its
benefit is the low fan-in per replica.  In this section, we devise a
different approach that sacrifices both the uniformity and the fan-in
in order to optimize the delay. We start with a specific
instance of our approach, called {\em Tree-Random}.
Tree-Random is a special case of a family of algorithms
$\ell$-Tree-Random, which we introduce later.
It is presented first to demonstrate one extremum, in terms of
its fan-in and delay, contrasting the Random algorithm.

We define the {\em Tree-Random} algorithm as follows.  We partition
the replicas into blocks of size $4t$, and arrange these blocks on the
nodes of a binary tree.  For each replica there are three interesting
sets of replicas.  The first set is the $4t$ replicas at the root of
the tree.  The second and third sets are the $4t$ replicas at the
right and left sons of the node that the replica is in.  The total
number of interesting replicas for each replica is at most $12t$, and
we call it the {\em candidate set} of the replica.  In each round,
each replica chooses $\Fout$ replicas from its candidate set uniformly
at random and sends a message to those replicas.

\begin{theorem}
\label{prop:eptree}
The delay of the Tree-Random algorithm is
$O(\frac{\coupona{\alpha}}{\Fout}+\frac{\log(\alpha)}{\Fout} +
\frac{t}{\Fout}\log(n/t))$ for $n > 8t$.
\end{theorem}

\begin{proof}
Let $u$ be any update.
We say that a node in the tree is active for $u$ if $2t$ correct replicas
(out of the $4t$ replicas in the node) are active for $u$.
We start by bounding the expected number of rounds, starting from $\eta_u$,
for the root to become active.
The time until the root is active can be bounded by the delay
of the Random algorithm with $4t+\alpha$ replicas.
Since on average one of every three messages is
targeted at the root,
within expected
$O(\coupona{\alpha}/\Fout+\log(\alpha)/\Fout)$
rounds the root becomes active.

The next step of the proof is to bound how much time it takes from
when a node becomes active until its child becomes active.  We will
not be interested in the expected time, but rather focus on the time
until there is at least a constant probability that the child is
active, and show a bound of $O(t/\Fout)$ rounds.

Given that $2t$ correct replicas in the parent node are active,
each replica in the child node has an expectation of receiving
$\Fout/12$ updates from new replicas in every round.
Using a Chernoff bound, this implies that in $\ell=96t/\Fout$
rounds each replica
has a probability of $e^{-t}$ of not becoming active.  The
probability that the child node is not active
(i.e. less of $2t$ of its replicas are active)
after $\ell$ rounds is bounded
by $b = 3te^{-t} < 5/6$ for $t \ge 2$.

In order to bound the delay we consider the delay until a leaf node
becomes active.  We show that for each leaf node, with high
probability its delay is bounded by $O(t\log(n/t))$.  Each leaf node
has $\log(n/4t)$ nodes on the path leading from the root to it.
Partition the rounds into meta-rounds, each containing $\ell$ rounds.
For each meta-round there is a probability of at least $1-b$ that
another node on the path would become active.  This implies that in
$k$ meta-rounds,
we have an expected number of $(1-b)k$ active nodes on the path.
Therefore, the probability that we have less than $(1-b)k/2$ is at
most $e^{-(1-b)k/8}$.
We have $\log(n/4t)$ nodes on the the path, this gives the constraint that
$k\geq 2\log(n/4t)/(1-b)$. In addition we like the probability that there
exists a leaf node that does not become active to
be less than $(t/n)^2$, which holds for $k \geq 16\log(n/4t)/(1-b)$.
Consider $k=16\log(n/4t)/(1-b)$ meta rounds.
Since there are at
most $n/4t$ leaves in the tree, then with probability at least
$1-4t/n>1/2$ the number of meta-rounds is at most $k=O(\log(n/t))$.
Thus, the delay is $k\ell=O(t\log(n/t)/\Fout)$.
This implies that the total
expected delay is bounded by $O(\coupona{\alpha}/\Fout +
\log(\alpha)/\Fout + t\log(n/t)/\Fout)$.
\end{proof}

Two points about this theorem are worthy of noting.
First, we did not attempt to optimize for the best
constants. In fact, we note that
much of the constant factor in the Tree-Random propagation delay
can be eliminated if we modify the algorithm to propagate messages
deterministically down the tree (but continue selecting targets at
random from the root node).

Second, the Tree-Random algorithm gains its speed at the expense of a
large fan-in. The replicas at the root of the tree receive $O(n)$
messages in each round of the protocol, and therefore in practice,
constitute a centralized bottleneck. Theorem~\ref{prop:tradeoff}
shows that in our model there
is an inherent tradeoff between the fan-in and the delay.

The next theorem claims a bound on the fan-in of the Tree-Random algorithm.

\begin{theorem}
\label{prop:trfan-in}
The fan-in of the Tree-Random algorithm is $\Theta(n\Fout/t)$, for
$n=\Omega(\frac{t}{\Fout}\log n)$.
\end{theorem}

\begin{proof}
Any replica at the root has a probability of $\Fout/(12t)$ receiving a
message from any other replica.  This implies that the expected number
of messages per round is $n\Fout/(12t)$, which establishes the lower
bound. The probability that a replica receives more than
$2 \frac{\Fout n}{12t}$ is bounded by $e^{- \Fout n/3(12t)}$ (using the
Chernoff bound).  Since $n=\Omega(\frac{t}{\Fout}\log n)$,
the probability is bounded by $1/n^2$, and the
theorem follows.
\end{proof}

We now define and analyze the generalized $\ell$-Tree-Random method.
We partition the replicas into blocks of size $\ell$, and arrange
these blocks on the nodes of a binary tree.  As in the Tree-Random algorithm,
for each replica there are three interesting sets of replicas.  The
first set is the $\ell$ replicas at the root of the tree.  The second
and third sets are the $\ell$ replicas at the right and left sons of
the node that the replica is in. The total number of replicas in the
three sets is at most $3\ell$, and we call it the candidate set of the
replica.  In each round, each replica chooses $\Fout$ replicas from its
candidate set uniformly at random and sends a message to those replicas.

Note that the Tree-Random propagation is simply setting $\ell=4t$ and the
random propagation is simply setting $\ell=n$.

\begin{theorem}
The $\ell$-Tree-Random algorithm has delay
\[
O\left(\frac{\coupona{\alpha}}{\Fout}
                \left(\frac{\ell+\alpha}{\alpha}\right)^{1-1/t} +
\frac{\log(\ell+\alpha)}{\Fout}
+ \frac{t}{\Fout}\log(n/\ell)\right)
\]
and fan-in $\Theta(n\Fout/\ell)$, for $4t \leq \ell \leq n\Fout/\log n$.
\end{theorem}

\begin{proof}
The proof of the fan-in is identical to the one of the Tree-Random
algorithm.  We have $\ell$ replicas at the root. Each replica sends
to each replica at the root with probability $\Fout/3\ell$. Therefore the
expected number of updates to each replica in the root is $n\Fout/3\ell$,
which establishes the lower bound on fan-in. With probability
$e^{-n\Fout/3(3\ell)}\leq 1/n^2$, each replica receives less than $2n\Fout/3\ell$
updates in a round.

The proof on the delay has two parts. The first is computing the time
it takes to make all the replicas in the root active.  This can be
bounded by the delay of the Random algorithm with $\ell + \alpha$
replicas, and so is
$O\left(\frac{\coupona{\alpha}}{\Fout}(\frac{\ell+\alpha}{\alpha})^{1-1/t} +
\frac{\log(\ell+\alpha)}{\Fout}\right)$.

The second part is propagating on the tree.  This part is similar to
the Tree algorithm.  As before, in each node at each round, each
replica has a constant probability of receiving messages from
$\Theta(\Fout)$ new replica.
This implies that with some constant probability $1-b$ all
the replicas in a node are active after $O(t/\Fout)$ rounds.  The analysis
of the propagation to a leaf node is identical to before, and thus
this second stage takes $O(\log (n/\ell))$ meta-rounds and the total
delay on the second stage is $O(\frac{t}{\Fout}\log (n/\ell))$.
\end{proof}

\section{Discussion}
\label{sec:discussion}

Our results for the Random and $\ell$-Tree-Random
algorithms are summarized in Table~\ref{tab:results}.

\begin{table*}[t]
\begin{center}
\begin{tabular}{|l|l|l|}
\hline
Method & Fan-in & Delay ($\alpha \ge 2t$) \\
\hline
Random & $O(\Fout+\log{n})$ &
$O\left( \frac{t}{\Fout}
(\frac{n}{\alpha})^{(1-\frac{1}{3t})} + \frac{\log(n)}{\Fout} \right)$
 \\
\hline
$\ell$-Tree-Random & $\Theta(n\Fout/\ell)$ &
$O\left(\frac{t}{\Fout}
                \left(\frac{\ell+\alpha}{\alpha}\right)^{1-1/t} +
\frac{\log(\ell+\alpha)}{\Fout}
+ \frac{t}{\Fout}\log(n/\ell)\right)$ \\
\hline
\end{tabular}
\end{center}
\caption{Properties of diffusion methods}
\label{tab:results}
\end{table*}

Using the fan-in/delay bound of Theorem~\ref{prop:tradeoff}, we now
examine our diffusion methods. The Random algorithm has
$O(\log{n})$-amortized fan-in of $O(\Fout)$, yielding a product of
delay and amortized fan-in of $O\left(t
(\frac{n}{\alpha})^{(1-\frac{1}{3t})} + \log(n)\right)$ when $\alpha
\ge 2t$.  This is slightly inferior to the lower bound in the range of
$t$ for which the lower bound applies.  The Tree-Random method has
fan-in (and amortized fan-in) of $O(n\Fout/t)$ and delay
$O(\frac{\log(\alpha)}{\Fout} + \frac{t}{\Fout}\log(n/t))$ if $\alpha
\ge 2t$.  So, their product is $O(\frac{n\log(\alpha)}{t} +
n\log(n/t))$, which again is inferior to the lower bound of
$\Omega(tn/\alpha)$ since $t/\alpha\leq 1$. However, recall from
Theorem~\ref{prop:lbdelay} that the delay is always
$\Omega(\frac{t}{\Fout}\log(\frac{n}{\alpha}))$, and so for the fan-in
of $O(n\Fout/t)$ it is impossible to achieve optimal delay/fan-in
tradeoff.
In the general $\ell$-Tree-Random method,
putting $\ell \ge \alpha \log(n/\alpha)$,
the $\ell$-Tree algorithm exhibits a fan-in/delay product of
at most $O(\frac{tn}{\alpha})$, which is optimal. If $\ell < \alpha
\log(n/\alpha)$, the product is within a logarithmic factor from
optimal. Hence, Tree propagation provides a spectrum of protocols that
have optimal delay/fan-in tradeoff to within a logarithmic factor.

Our lower bound of $\Omega(\frac{t}{\Fout}\log{\frac{n}{\alpha}})$ on
the delay of any diffusion protocol says that we pay a high price for
Byzantine fault tolerance: when $t$ is large, diffusion in our model
is (necessarily) slower than diffusion in system models admitting only
benign failures.  By comparison, in systems admitting only benign
failures there are known algorithms for diffusing updates with $O(\log
n)$ delay, including one on which the Random algorithm studied here is
based~\cite{Pit87}.

\section{Simulation Results}
\label{sec:simulation}

\begin{figure*}[t]
\begin{center}
\begin{tabular}{cc}
{\epsffile{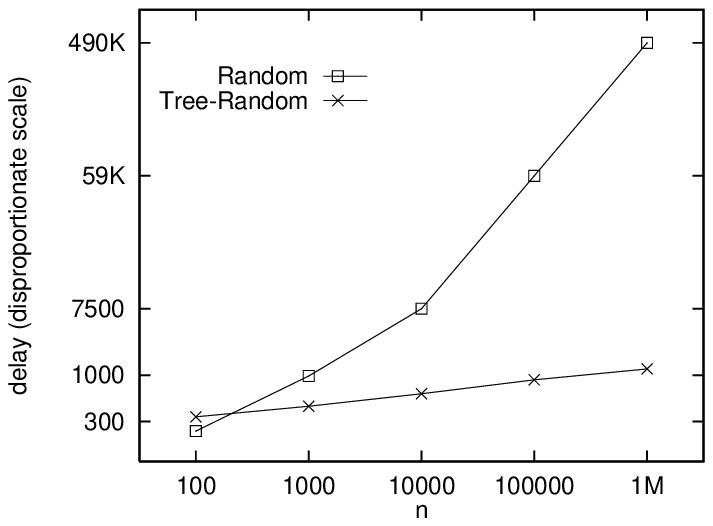}} & {\epsffile{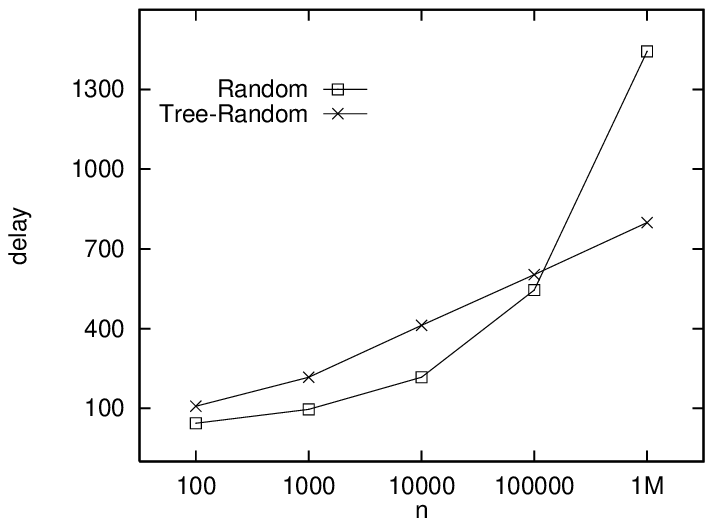}} \\
(a) $t = 16$, $\alpha = 17$ & (b) $t = 16$, $\alpha = \sqrt{2tn}$
\end{tabular}
\end{center}
\caption{Delay of Random and Tree-Random algorithms.}
\label{fig:sim}
\end{figure*}

Figure~\ref{fig:sim} depicts simulation results of the Random and
Tree-Random algorithms. The figure portrays the delay of the two
methods for varying system sizes (on a logarithmic scale), where $t$
was fixed to be $16$.  In part (a) of this figure, we took the size
$\alpha$ of the initial set $I_u$ to be $\alpha=t+1$.  This graph
clearly demonstrates the benefit of the Tree-Random method in these
settings, especially for large system sizes.  In fact, we had to draw
the upper half of the $y$-axis scale in this graph disproportionately
in order for the
small delay numbers of Tree-Random, compared with the large delay
numbers exhibited by the Random method, to be visible.  Part (b) of
the graph uses $\alpha=\sqrt{2tn}$, which reflects the minimal initial
set that we would use in the Fleet system, which is one of the
primary motivations for our study (see Section~\ref{sec:motivation}).
For such large initial sets, Random
outperforms Tree-Random for all feasible systems sizes, and the
benefit of Tree-Random is only of theoretical interest (e.g., $n >
1000000$).

In realistic large area networks, it is unlikely that $100\%$
of the messages arrive in the round they were sent in, even for a fairly
large inter-round period. In addition, it may be desirable to set the
inter-round delay reasonably low, at the expense of letting some
messages arrive late. Some messages may be dropped in realistic
scenarios, and hence, to accommodate such failures, we also ran our
simulations while relaxing our synchrony assumptions. In these
simulations, we allowed some threshold---up to $5\%$---of the messages
to arrive in later rounds than the rounds they were sent or to be
omitted by the receiver. The resulting behavior of the protocols were
comparable to the synchronous settings. We conclude that our protocols
can just as effectively be used in asynchronous environments in which
the inter-round delay is appropriately tuned.

\section{Conclusion}
\label{sec:conclusion}

In this paper we have provided the first analysis of epidemic-style
update diffusion in systems that may suffer Byzantine component
failures.  We require that no spurious updates be accepted by correct
replicas, and thus that each correct replica receive an update from
$t$ other replicas before accepting it, where the number of faulty
replicas is less than $t$.  In this setting, we analyzed the delay and
fan-in of diffusion protocols.  We proved a lower bound on the delay
of any diffusion protocol, and a general tradeoff between the delay
and fan-in of any diffusion protocol.  We also proposed two diffusion
protocols and analyzed their delay and fan-in.

\bibliographystyle{latex8}

\end{document}